\documentclass[conference]{IEEEtran}
\IEEEoverridecommandlockouts
\usepackage{cite}
\usepackage{amsmath,amssymb,amsfonts}
\usepackage{algorithmic}
\usepackage{graphicx}
\usepackage{textcomp}
\usepackage{xcolor}

\usepackage{capt-of}
\usepackage{booktabs}
\usepackage{multirow}

\usepackage{natbib}
\setcitestyle{numbers,square}

\def\BibTeX{{\rm B\kern-.05em{\sc i\kern-.025em b}\kern-.08em
    T\kern-.1667em\lower.7ex\hbox{E}\kern-.125emX}}
\begin{document}

\title{TSNet-SAC: Leveraging Transformers for Efficient Task Scheduling}

\author{

\IEEEauthorblockN{1\textsuperscript{st} Ke Deng, 2\textsuperscript{nd} Zhiyuan He, 3\textsuperscript{rd} Hao Zhang, 4\textsuperscript{th} Haohan Lin, 5\textsuperscript{th} Desheng Wang}
\IEEEauthorblockA{\textit{Wuhan National Laboratory for Optoelectronics} \\
\textit{Huazhong University of Science and Technology}\\
Wuhan, China \\
sheepsui\@hust.edu.cn, zedyuanhe\@hust.edu.cn, haoz52\@hust.edu.cn, M202272477\@hust.edu.cn, dswang\@hust.edu.cn}}

\maketitle

\begin{abstract}
In future 6G Mobile Edge Computing (MEC), autopilot systems require the capability of processing multimodal data with strong interdependencies. However, traditional heuristic algorithms are inadequate for real-time scheduling due to their requirement for multiple iterations to derive the optimal scheme. We propose a novel TSNet-SAC based on Transformer, that utilizes heuristic algorithms solely to guide the training of TSNet. Additionally, a Sliding Augment Component (SAC) is introduced to enhance the robustness and resolve algorithm defects. Furthermore, the Extender component is designed to handle multi-scale training data and provide network scalability, enabling TSNet to adapt to different access scenarios. Simulation demonstrates that TSNet-SAC outperforms existing networks in accuracy and robustness, achieving superior scheduling-making latency compared to heuristic algorithms. 
\end{abstract}

\begin{IEEEkeywords}
Transformer, 6G, autopilot, task offloading, mobile edge computing
\end{IEEEkeywords}

\section{Introduction}
The MEC architecture has been introduced as one of the enabling technologies in 6G to address the stringent requirements of data-intensive applications, especially autopilot, which demand high computational performance and low service latency. The autopilot requires processing massive sensor data, such as LiDAR, cameras, radar, and ultrasonic sensors, to perceive and understand the surrounding environment \cite{h01}. These sensors produce a colossal amount of multimodal data with solid interdependency. For example, LiDAR is mainly used to obtain the vehicle's surrounding three-dimensional point cloud data, while the camera is for image data. The data from these two sensors are interdependent, i.e., the point cloud data supplements the image to improve the system's perception accuracy of the environment, as the image help LiDAR recognize and locate the target objects\cite{h02,h03}. Therefore, in MEC, a more extensive task granularity guarantees the computing offloading to avoid the impact of frequent data transmission and processing on system performance. 

However, due to the complexity of road conditions, sensor data usually has ambiguity\cite{h04}. Although traditional heuristic algorithms can adapt to diversified variable problems and provide appropriate scheduling, they require multiple iterations failing to meet the high requirements for the delay of safe autopilot \cite{h05}. Despite a feasible method of using neural networks for real-time processing and analysis of sensor data, such as Multi-Layer
Perceptron (MLP)\cite{h06}, is far from proper scheduling due to their limited learning capacity in the coupling offloading tasks.

Fortunately, the breakthroughs made in Transformer recently have brought us new ideas. It \cite{b25} has achieved SOTA results in machine translation tasks. Due to the agility of self-attention in coping with the correlation data, Transformer has attracted wide attention from academia and industry. It has been widely used in Natural Language Processing (NLP) \cite{b26,b27} and other fields, including visual \cite{b28,b29} and audio processing \cite{b30,b31}. Additionally, MLP-Mixer \cite{b37} replaces the self-attention structure with information interaction by feature fusion, simplifies the network parameters, and achieves similar performance to Transformer.

Based on the above structure for feature extraction, we innovatively design the TSNet-SAC to make strategies in task offloading and resource allocation. Simulation shows the great potential of this network in the deployment of MEC. The main contributions are as follows:

\begin{itemize}
    \item This paper models the deployment of MEC with a single base station and multiple terminals. Moreover, numerous optimized strategies are generated based on this model by applying Genetic Algorithm (GA).
    \item This paper innovatively proposes a TSNet based on Transformer to tackle the task offloading and resource allocation efficiently. 
    \item Sliding Augment Components (SAC) and Extenders were developed to improve the performance of the TSNet, resulting in increased accuracy and enhanced scalability and robustness.
\end{itemize}

The rest of this paper is organized as follows. In section \ref{System Model and Problem Formulation}, we define the system model and propose the scheduling optimization problem. Then, we propose TSNet-SAC to solve the problem, which is explained and discussed in depth in section \ref{Problem Solution}. Section \ref{Evaluation} verifies the feasibility and effectiveness of this method through a series of simulations. Finally, some conclusions are drawn in Section \ref{Conclusion}.

\section{System Model and Problem Formulation}\label{System Model and Problem Formulation}
The single-core MEC with a multi-accessed single base station is modeled based on the reference\cite{h07} and the task offloading is formulated as $\{0, 1\}$ integer programming. In contrast, resource allocation, including computing resources, channel bandwidth, and transmission power, is abstracted as continuous variable programming. The model contains $N$ terminals and $1$ MEC server. Each terminal generates a task within the unit time interval whose information is represented by the quad $\{u_i, c_i, d_i, h_i\}$.

\begin{itemize}
    \item $u_i$ represents the amount of data in the uplink; 
    \item $c_i$ represents the CPU cycles required by the terminal or MEC server to compute the task; 
    \item $d_i$ represents the amount of data in the downlink; 
    \item $h_i$ represents the transmission channel parameters from the terminal to the server.
\end{itemize}

The task offloading of terminals is represented by the binary $m_i$. When $m_i=0$, it means that the task is computed locally; $m_i=1$ means that the service is offloaded to the edge server for computing.

\subsection{Task Computing}\label{Task Execution Model}

In the following sections, we will outline delay and energy consumption for tasks offloading or terminal computing.

\subsubsection{Terminal Computing}\label{Terminal Computing}

The computing delay $T_i^{loc}$ of the task is computed in terminals can be expressed as:

\begin{align}
    T_i^{loc}=(1-m_i)c_i/f^{loc}
    \label{eq1}
\end{align}

Where $f^{loc}$ is the terminal CPU computing frequency. Also, the computing energy consumption $E_i^{loc}$ can be expressed as:

\begin{align}
    E_i^{loc} = k^{loc}(f^{loc})^2(1-m_i)c_i
    \label{eq2}
\end{align}

Where $k^{loc}$ is a constant related to the type of terminal CPU.

\subsubsection{Offloading}\label{Offloading}

According to Shannon's Formula, the transmission delay $T_i^{ul}$ and $T_i^{dl}$ of uplink and downlink can be separately expressed as:  

\begin{align}
    T_i^{ul}=m_iu_i/(W_i^{ul}{log}_2{(}1+\frac{p_i^{ul}h_i^{ul}}{N_0W^{ul}}))
    \label{eq3}
\end{align}

\begin{align}
    T_i^{dl}=m_id_i/(W_i^{dl}{log}_2{(}1+\frac{p_i^{dl}h_i^{dl}}{N_0W^{dl}}))
    \label{eq4}
\end{align}

Where $h_i^{ul}$ and $h_i^{dl}$  are the channel parameters, $N_0$ represents the power spectral density of Gaussian White Noise in the channel. $W_i^{ul}$ and $W_i^{dl}$ are the bandwidth of uplink and downlink while the transmission power is $p_i^{ul}$ and $p_i^{dl}$ separately. Similarly, the transmission energy consumption $E_i^{ul}$ and $E_i^{dl}$ can be expressed as: 

\begin{align}
    E_i^{ul}=m_iu_ip_i^{ul}/(W_i^{ul}{log}_2{(}1+\frac{p_i^{ul}h_i^{ul}}{N_0W^{ul}}))
    \label{eq5}
\end{align}

\begin{align}
    E_i^{dl}=m_id_ip_i^{dl}/(W_i^{dl}{log}_2{(}1+\frac{p_i^{dl}h_i^{dl}}{N_0W^{dl}}))
    \label{eq6}
\end{align}

Thus, the delay and energy consumption of the offloading tasks can be expressed as:

\begin{align}
    T_i^{exe}=m_ic_i/f_i^{AP}
    \label{eq7}
\end{align}

\begin{align}
    E_i^{exe}=k^{AP}(f_i^{AP})^2m_ic_i
    \label{eq8}
\end{align}

Where $k^{AP}$ is a power constant related to the type of the MEC CPU and $f^{AP}$ is its computing frequency. Finally, the total delay of terminal $i$ is:

\begin{align}
  T_i =
  \begin{cases}
    T_i^{loc}, & m_i=0 \\
    T_i^{exe}+T_i^{ul}+T_i^{dl}, & m_i=1
  \end{cases}
  \label{eq9}
\end{align}

\subsection{Problem Formulation}\label{Problem Formulation}

It can be given from section \ref{Task Execution Model} that the system delay is defined as $T=\frac{1}{N}\sum_{i=1}^{N}T_i$ and the total system energy consumption is denoted as $E=\sum_{i=1}^{N}{(E_i^{dl}+E_i^{exe}+E_i^{ul}+E_i^{loc})}$. According to \cite{h08}, the optimization of them is proved to be a mixed integer nonlinear programming (MINLP) problem. To simplify it, we consider establishing a utility index to couple the multi-objective into one, which is defined as $U=\lambda T+(1-\lambda)E$, where $\lambda\in[0,1]$ is the system preference coefficient. The optimization of the model can be established as \eqref{eq10}: 

\begin{align}
	\underset{m_i,p_i^{ul},p_i^{dl},f_i^{AP}}{\mathrm{minimize}}  \quad& U \label{eq10}\\
	\notag
	\text{subject to:}
	\quad& f_{min}^{AP} \leq f_{i}^{AP} \leq f_{max}^{AP} \tag{10a}\label{eq10:con2}\\
	\quad& p_{min}^{ul} \leq p_{i}^{ul} \leq p_{max}^{ul} \tag{10b}\label{eq10:con3}\\
	\quad& p_{min}^{dl} \leq p_{i}^{dl} \leq p_{max}^{dl} \tag{10c}\label{eq10:con4}\\
        \quad& \sum^N_{i=1} f^{AP}_i \leq F^{AP}_{total} \tag{10d}\label{eq10:con5}\\
        \quad& m_i \in \{0,1\}, \forall i \in \{1,2,\ldots ,N\} \tag{10e}\label{eq10:con6}
\end{align}

Among them, \eqref{eq10:con2} denotes the computational frequency constraint where $f_{i}^{AP}$ is the transmission frequency, \eqref{eq10:con3} and \eqref{eq10:con4} represents the transmit power constraint of the mobile terminal and the edge server, respectively, and \eqref{eq10:con5} denotes the total amount of computing frequency constraint that the edge server can allocate, \eqref{eq10:con6} denotes task uniqueness constraints.

\section{Problem Solution}\label{Problem Solution}

\begin{figure*}
    \centerline{\includegraphics[width=0.95\textwidth]{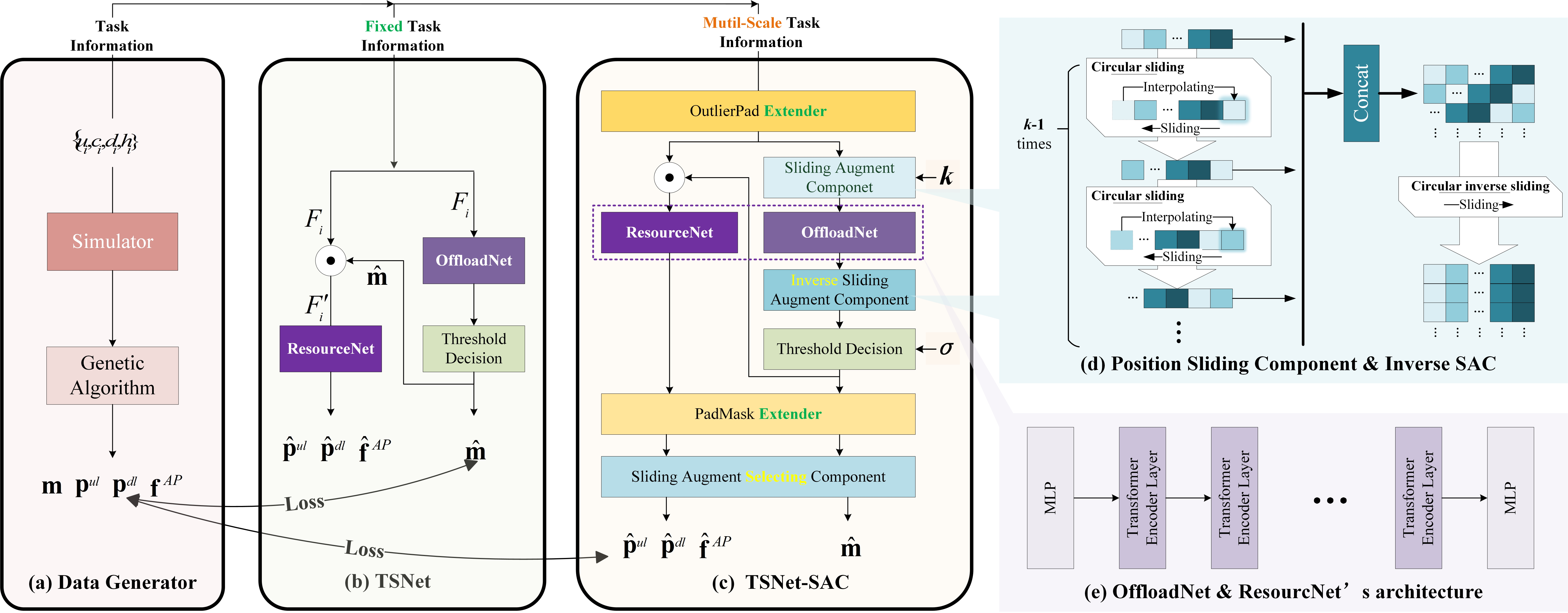}}
    \caption{Proposed TSNet-SAC. (a) Data Generator, (b) TSNet, (c) TSNet-SAC, (d) Sliding Augment Component \& Inverse Sliding  Augment Component, (e) OffloadNet \& ResourceNet' s Architecture.}
    \label{fig:1}
\end{figure*}

\subsection{TSNet}\label{TSNet}

To address the unacceptable time cost of multiple iterations in the traditional optimization algorithm, we propose the TSNet, which employs Transformer architecture to reduce optimizing latency for real-time scheduling. We utilize the GA to generate strategies for training, containing task information with labels of offloading decisions and resource allocations\cite{h09}, as shown in Fig.~\ref{fig:1} (a).

Given the constraint of limited system resources, task offloading decisions for the current and other terminals are intricately interrelated. Therefore, in designing networks, it is necessary to extract not only the specific task features but also the potential relationship among different tasks.

It’s noticed that in NLP\cite{b26,b27}, the Transformer has achieved great success in extracting the feature information of individual words, as well as capturing the potential correlation features in a sentence. As the inter-correlation among words is proximate to the resource-constrained conditions of various terminal tasks in scheduling, we leverage the Transformer architecture to design the TSNet to make the task offloading and resource allocation, as shown in Fig.~\ref{fig:1} (b). 

TSNet comprises two sub-networks: OffloadNet, which focuses on task offloading, and ResourceNet, which dedicates resource allocation. Through extracting and integrating potential interdependent features between tasks, these sub-networks work in tandem to make an informed scheduling strategy. 

OffloadNet employs a three-stage architecture in Fig.~\ref{fig:1} (e), which begins with the MLP layer to extract High-Dimensional Features by embedding input data $F_i$ to $F_e$, whose dimension from $(B \times N \times 4)$ to $(B \times N \times 32)$. Subsequently, a multiple Transformer Encoder Layer is leveraged to make feature fusion, the output is  $F_t$ with a dimension of $(B \times N \times 32)$. Finally, a binary classification process is conducted by an MLP on the fused feature $F_t$, to generate the offloading decision $\hat{\boldsymbol{m}}$, with dimension $(B \times N)$. 

The other sub-network, ResourceNet, differs from OffloadNet in that its output is a resource allocation strategy $\boldsymbol{p_i^{ul},p_i^{dl},f_i^{AP}}$, represented as a $(B\times N\times 3)$ tensor, generated by the MLP layer at the end. Apart from this, the network structure of the two sub-networks is similar.

To cope with the large network depth by joint training of the two sub-networks which cause gradient vanishing, we train them separately to reduce complexity. On the other hand, there is a strong coupling between offloading strategy and resource allocation in task scheduling. Therefore, we use the dot product between the offloading strategy $\hat{\boldsymbol{m}}$ output by OffloadNet and the task information $F_i$ to obtain $F_i^{\prime}$, which is used as input to ResourceNet. In all, this approach establishes the interrelation between OffloadNet and ResourceNet. The dot product can be expressed as:

\begin{align}
    F_i^{\prime}=F_i\odot\boldsymbol{\hat{m}}
    \label{eq11}
\end{align}

By combining OffloadNet and ResourceNet, TSNet is able to achieve real-time task scheduling with fixed access counts.

\subsection{Extenders}\label{TSNet}

The input of fixed access counts causes TSNet to be trained for specific access scenarios, limiting its predictive capabilities and incompatible with the different accessed physically. To make the TSNet adapt to the different access counts, we designed the Extender component to unify the input data to a consistent dimension.

As shown in Fig.~\ref{fig:1} (c), the Extender component comprises two parts: OutlierPad Extender and PadMask Extender. OutlierPad Extender is located at the beginning of TSNet and expands the specific access counts, $N$, to the upper limit of terminal access, $\Bar{N}$, thereby increasing the adaptability of TSNet to varying scenarios from a training dimension of $(B \times N \times 4)$ to $(B \times \Bar{N} \times 4)$. PadMask Extender, located at the end of the network model, masks the filled positions by OutlierPad Extender, outputting task offloading decisions corresponding to the original access count. The joint effect of OutlierPad Extender and PadMask Extender enables TSNet to adapt to multi-scale data training, adapt to diverse access scenarios, and improve the accuracy and scalability of the network.

\subsection{Sliding Augment Component}\label{Sliding Augment Component}

Three critical challenges remain after designing the network architecture and Extender components, including model adaptation, insufficient quality of GA strategies, and outliers of offloading decisions.

Specifically, in a resource-constrained environment, the mutual constraint should be taken into consideration when scheduling independent tasks. In contrast, a normal Transformer utilizes absolute position encoding, which contains the sequence affecting the attention levels. To mitigate the position-specific relationship without sequential inherent, we propose the Sliding Augment Component (SAC), which applies $k-1$ times circular shift to the combination of multiple tasks, ensuring the equal probability of them appearing at different positions of the Transformer, which realized the adaptation of Transformer to the positional correlation between tasks, as shown in Fig.~\ref{fig:1} (d).

Furthermore, the increasing access count leads to the expansion of the problem scale, and the quality of strategies generated by GA typically deteriorates gradually \cite{h09}. To address this issue, the combination of restructured task information generated by the SAC is fed into OffloadNet. This expands the original single strategy to $k$ candidate, mitigating the reduction of strategy quality caused by the expansion of network scale in GA.

Moreover, due to the black-box nature of the neural network, the resource allocation by ResourceNet, including $\boldsymbol{p_i^{ul},p_i^{dl},f_i^{AP}}$, may exceed the constraint. By allocating the upper and lower limits of the constraint to the values that exceed the constraint \eqref{eq10:con2}\eqref{eq10:con3}\eqref{eq10:con4}\eqref{eq10:con5}, the resource allocation is constrained within a valid range. Furthermore, the utility index \eqref{eq10} is utilized to select candidate solutions to obtain optimal decisions.

\section{Evaluation}\label{Evaluation}

The simulation platform utilized in this study was a server equipped with an i9 CPU, four NVIDIA GeForce RTX 3090 GPUs, 20GB DDR RAM, and 5.4TB SSD. 

\subsection{Simulation Settings}\label{Simulation Settings}

The model parameter settings are shown in Table \ref{tab2}.

\begin{table}
    \caption{parameter settings for the simulation}
    \renewcommand{\arraystretch}{1.5}
    \label{tab2}
    \begin{center}
	\begin{tabular}{|c|p{4.5cm}|c|}
		\hline
		\textbf{Symbol}  & \textbf{Parameter}   & \textbf{Value} \\
		\hline
		$f^{loc}$         & Terminal CPU frequency                               & 2GHz \\
        \hline
		$k^{loc}$         & Constant related to terminal CPU               & $3 \times 10^{-27}$ \\
		\hline
        $k^{AP}$          & Constant related to MEC CPU                   & $1 \times 10^{-27}$ \\
		\hline
        $p^{ul}_{min}$    & Minimum transmission power of terminal                & 50mW \\
	  \hline	
        $p^{ul}_{max}$    & Maximum transmission power of terminal                & 200mW \\
		\hline
        $p^{AP}_{min}$    & Minimum transmission power of MEC                     & 20W \\
		\hline
        $p^{AP}_{max}$    & Maximum transmission power of MEC                     & 200W \\
		\hline
        $f^{AP}_{min}$    & Minimum assignable computing frequency of terminal    & 1GHz \\
		\hline
        $f^{AP}_{max}$    & Maximum assignable computing frequency of terminal    & 8GHz \\
		\hline
        $F^{AP}_{total}$  & Total assignable computing frequency of MEC           & 140GHz \\
		\hline
        $N_0$             & Power spectral density of channel noise               & -173dBm/Hz \\
		\hline
        $\lambda$         & Preference coefficient between delay and energy        & 0.5 \\
		\hline
        $\overline{N}$    & Maximum number of accesses                            & 40 \\
		\hline
	\end{tabular}
    \end{center}
\end{table}

\subsection{Competence of the Network}\label{Competence of the Network}

The simulation compares the learning capabilities of OffloadNet and ResourceNet. Specifically, the task offloading and resource allocation strategies were evaluated using the metrics of prediction accuracy and mean square error (MSE), respectively. To establish baselines for comparison, MLP and MLP-Mixer were adopted as baselines. The results of the experiments are presented as follows:

\begin{figure}
    \centering
    \centerline{\includegraphics[width=0.5\textwidth]{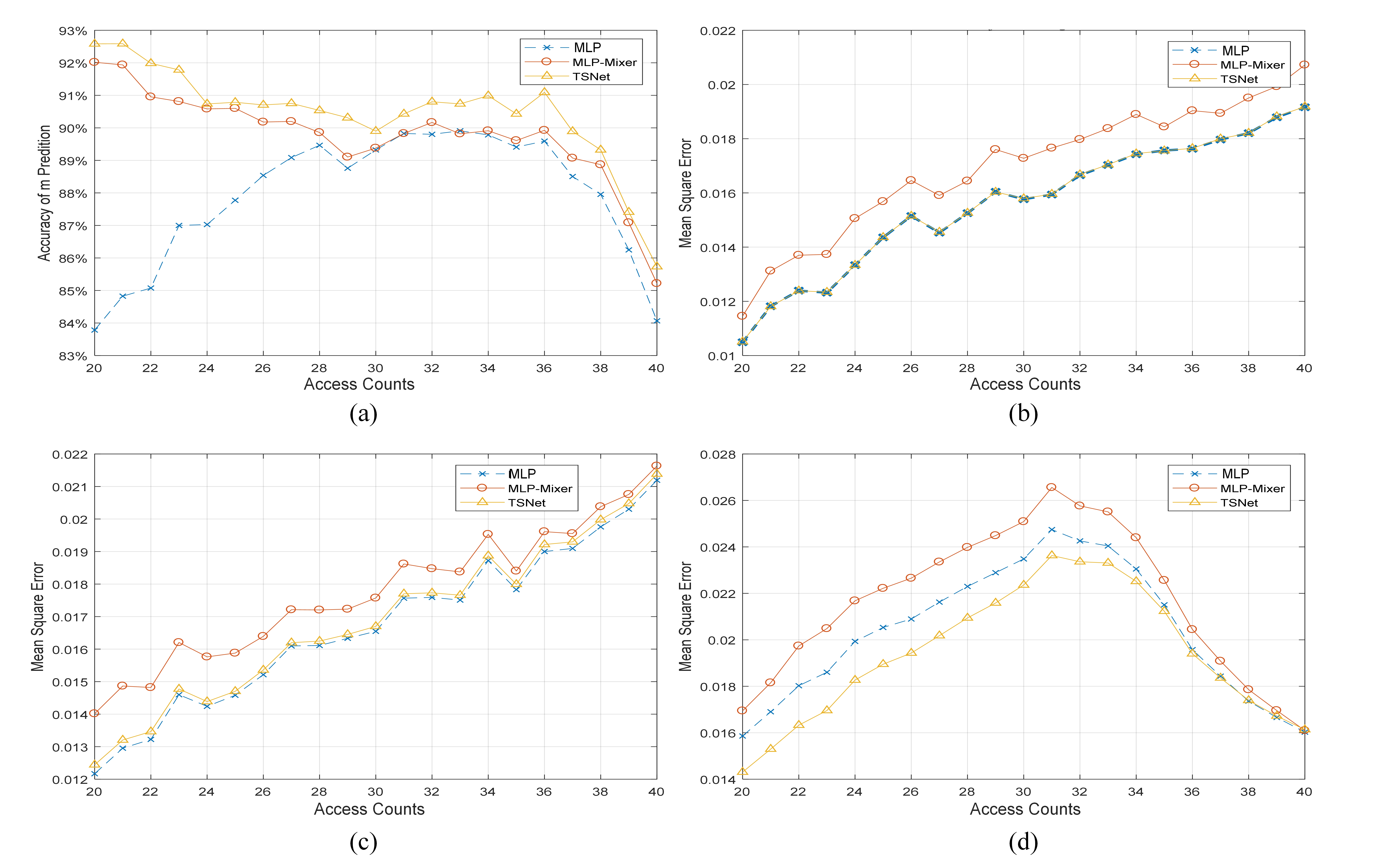}}
    \caption{Performance of sub-networks. (a) OffloadNet vs baselines. (b) ResourceNet vs baselines in predicting $p_i^{ul}$. (c) ResourceNet vs baselines in predicting $p_i^{dl}$. (d) ResourceNet vs baselines in predicting $f_i^{AP}$.}
    \label{fig:2}
\end{figure}

\begin{table*}[htbp]
  \centering
  \begin{minipage}[b]{0.28\linewidth}
    \centering
    \centerline{\includegraphics[width=\textwidth]{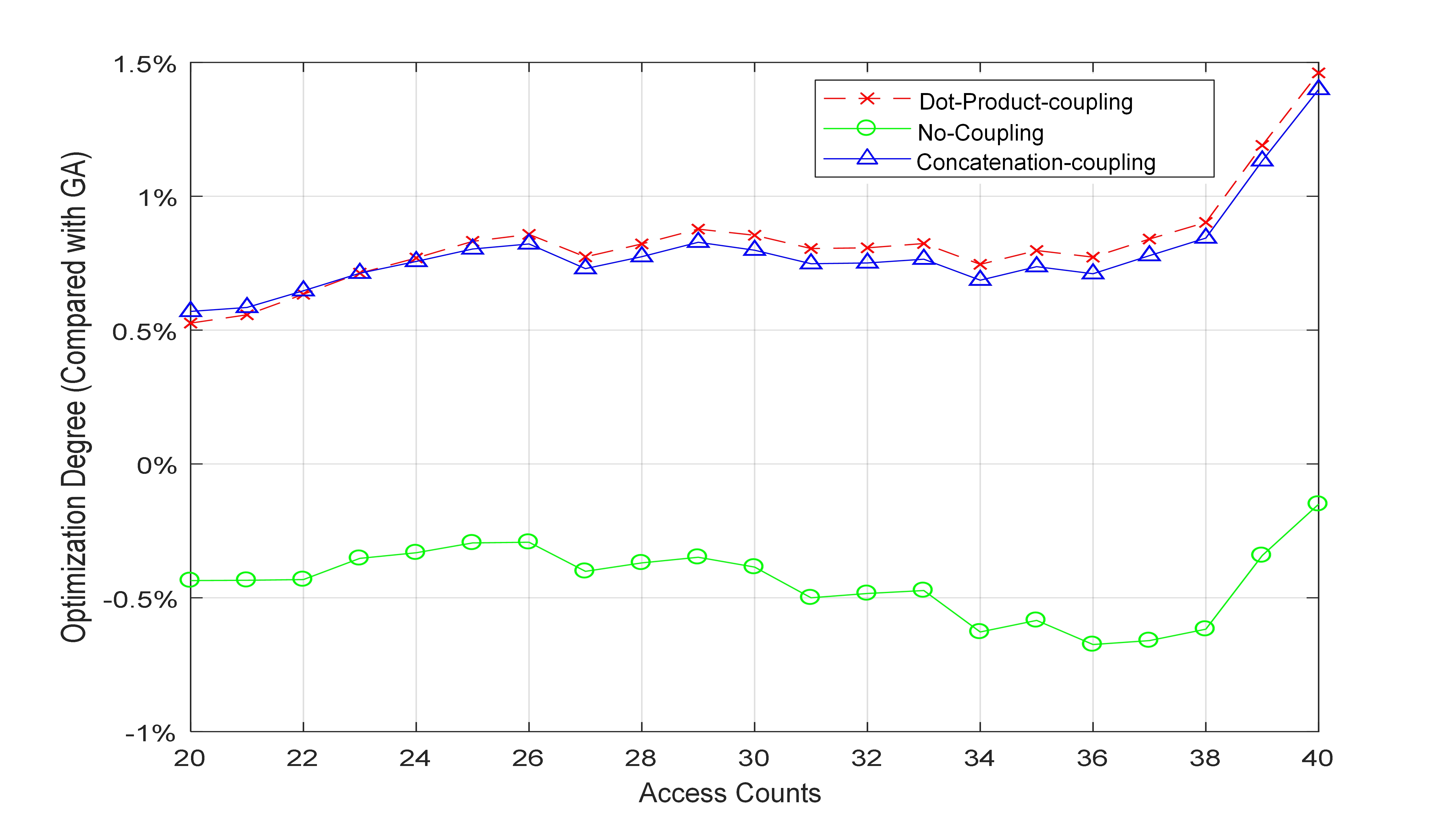}}
    \captionof{figure}{Effect of different coupling on the performance of the proposed TSNet.}
    \label{fig:3}
  \end{minipage}
    \hfill
    \begin{minipage}[b]{0.68\linewidth}
    \centering
    \caption{Performance comparison of different models}
    \label{tab1}
    \begin{tabular}{lcccccccc}
        \toprule
        \multirow{2}{*}{Method} & \multicolumn{8}{c}{Terminal count} \\
        & 20 & 23 & 26 & 29 & 32 & 35 & 38 & 40 \\
        \midrule
        GA & 0.16314 & 0.16054 & 0.15767 & 0.15412 & 0.15055 & 0.14665 & 0.14248 & 0.14086 \\
        MLP & 0.16475 & 0.16082 & 0.15698 & 0.15303 & 0.14951 & 0.14588 & 0.14192 & 0.13999 \\
        MLP-Mixer & \textbf{0.16226} & \textbf{0.15938} & 0.15633 & 0.15282 & 0.14943 & 0.14561 & 0.14130 & 0.13885 \\
        TSNet & 0.16228 & 0.15940 & \textbf{0.15632} & \textbf{0.15277} & \textbf{0.14934} & \textbf{0.14548} & \textbf{0.14120} & \textbf{0.13880} \\
        \bottomrule
    \end{tabular}
    \vspace{\baselineskip}
    \end{minipage}
  
\end{table*}

Fig.~\ref{fig:2} (a) shows that OffloadNet outperforms the baselines in predicting offloading decisions, achieving a prediction accuracy of $92.6\%$ in 20-terminal access scenarios and $86\%$ in 40-terminal access scenarios. In comparison to MLP and MLP-Mixer, ResourceNet shows a similar superiority to OffloadNet in Fig.~\ref{fig:2} (b-d). Specifically, deteriorating performance is observed in both ResourceNet and the baseline when the number of access terminals is more than 30. The reason is, as the problem scales up, the optimal resource allocation strategy of the model tends to favor full power load, resulting in decreased difficulty in predicting $\boldsymbol{f^{AP}}$ and reduced MSE in the prediction results. Overall, ResourceNet maintains a performance advantage despite the observed trend shift.

In general, the results demonstrate that OffloadNet and ResourceNet are capable of learning the features of offloading decisions and resource allocation. Next, we couple OffloadNet and ResourceNet during the prediction phase to form the integrated network TSNet. To achieve the best performance of the network, we investigate the effects of three coupling strategies, namely Dot-product-coupling, Concatenation-coupling, and No-coupling, on the performance of TSNet as shown in Fig.~\ref{fig:3}.


The simulation demonstrates a significant superiority of coupling networks to No-coupling, with Dot-product-coupling outperforming Concatenation-coupling to some extent. After constructing TSNet using dot product coupling, we compared the prediction results of TSNet with the baselines, using the average value of the utility metric as the evaluation criterion. The results are presented in Table \ref{tab1}.

It can be observed that TSNet outperforms baselines overall. Therefore, we can conclude that TSNet has better robustness and is more suitable for task offloading and resource allocation in MEC deployment, especially for cases with fixed terminal access count. 


\subsection{Extender Evaluation}\label{Extender Evaluation}

\begin{figure}[htbp]
    \centering
    \centerline{\includegraphics[width=0.5\textwidth]{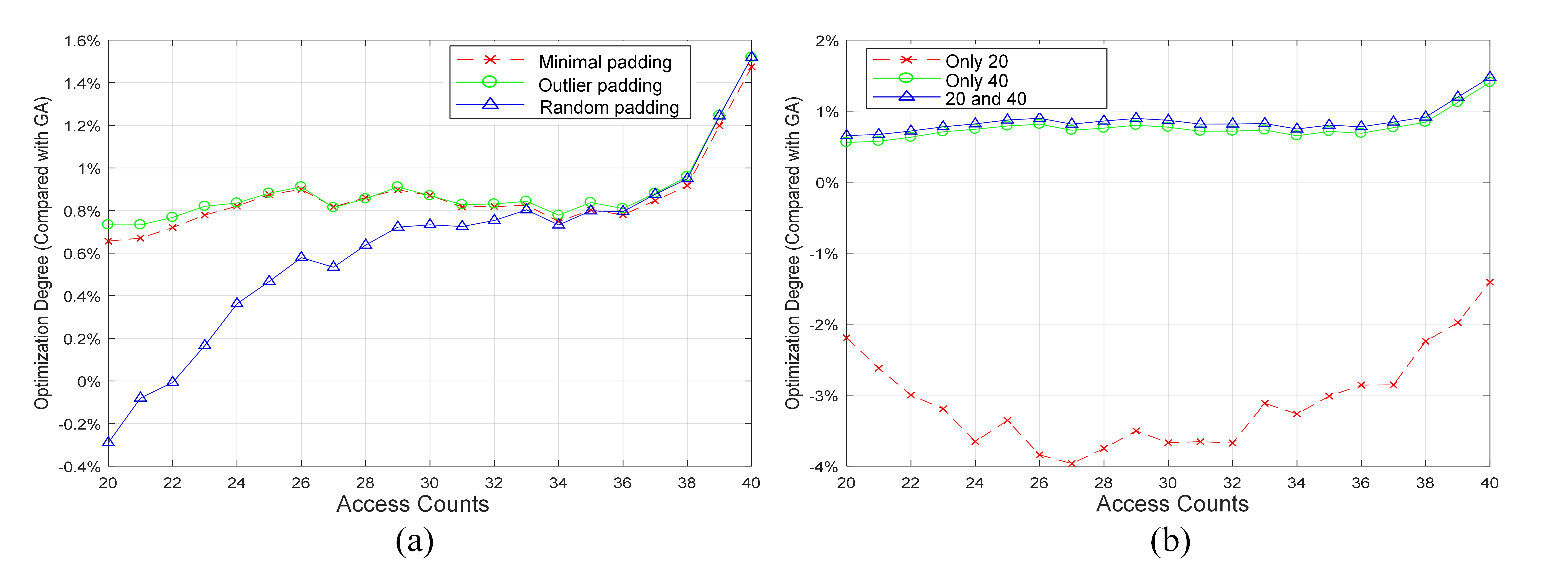}}
    \caption{Effect of regulating the Extender component on the performance of the proposed TSNet. (a) Different padding. (b) Different trainset compositions.}
    \label{fig:4}
\end{figure}

The performance and scalability of the network can be improved by regulating the Extender component. The simulation was conducted to investigate the effect of different padding and trainset compositions with different terminal access counts on the network performance. The gain on the GA strategies was used as the evaluation metric, as shown in Fig.~\ref{fig:4}.

Fig.~\ref{fig:4} (a) compared three different padding in the TSNet, namely, random values within the range of 0-1, a minimum value of 0, and an outlier of -1, with task information normalized to the 0-1 range. The results indicated that the outlier padding outperformed the others when the terminal access count was less than 30, which was caused by the approach recording the heteroscedastic data as a token enabling the identification of the terminal access. When access counts are sufficient, the impact of the three padding techniques tends to be similar, as few positions require padding.

Fig.~\ref{fig:4} (b) illustrates a significant improvement in performance when the terminal access count is 40 compared with the 20 counts. Combining training data with 20 and 40 terminal connections further improved performance, highlighting the benefits of multi-scale data training. In summary, Extender enhances the network's scalability for different terminal access as well as performance.

\subsection{SAC Evaluation}\label{SAC Evaluation}

\subsubsection{Why We Design SAC}\label{Why We Design SAC}

It is interesting to note that we observed the predictions of TSNet to be superior to MLP-Mixer in most cases, and even outperforming GA-generated strategies. We attribute this to the neural network learning optimization paradigm.

\begin{figure}[htbp]
    \centering
    \centerline{\includegraphics[width=0.5\textwidth]{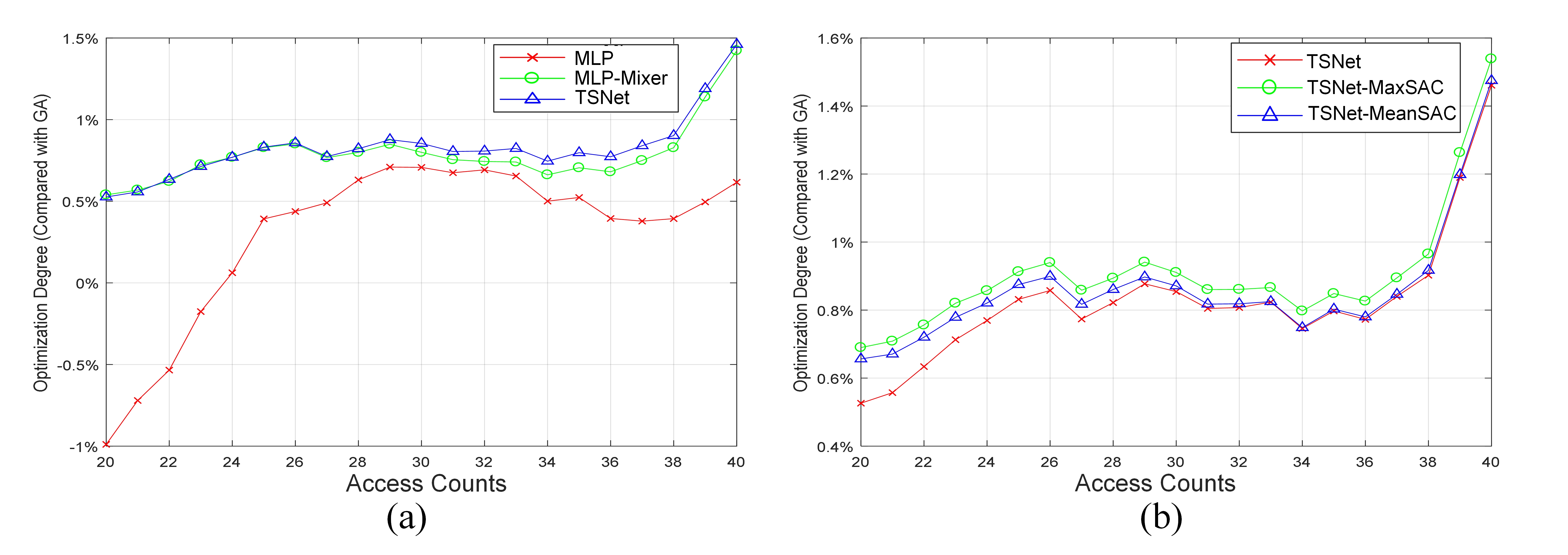}}
    \caption{Why did we design SAC? (a) Performance of TSNet comparing with baselines. (b) Effect of different methods for screening candidate solutions in SAC on the performance of the proposed TSNet-SAC.
}
    \label{fig:6}
\end{figure}

It is evident that the gain increases as the terminal access grows. However, the experimental observation contradicts that the network performance decreases with the problem scale \cite{h10}. Therefore, Fig.~\ref{fig:6} (a) validates the gradual deterioration of GA-generated optimization data with increasing problem scale.

To improve the performance of TSNet, we designed the SAC component, and Fig.~\ref{fig:6} (b) showed that the SAC structure improved the network performance in any terminal connection scenario. Compared to the original TSNet, TSNet-SAC increased the average optimization ratio of GA data by 1\%. Therefore, it can be concluded that the SAC component enhances the network's robustness to GA noise data.

\subsubsection{SAC Optimization}\label{SAC Optimization}

\begin{figure}[htbp]
    \centering
    \centerline{\includegraphics[width=0.5\textwidth]{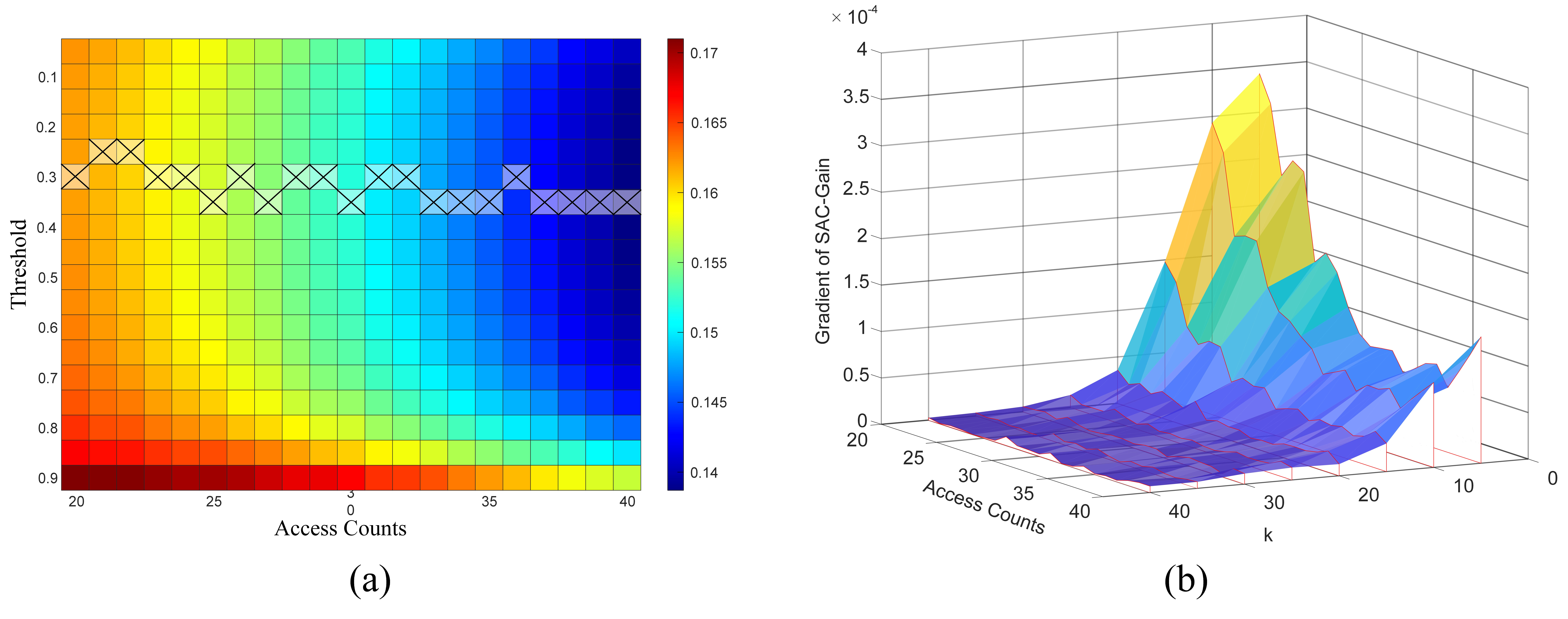}}
    \caption{Effect of different method for controlling SAC on the performance of the proposed TSNet-SAC. (a) The threshold in SAC (b) sliding hyperparameter $k$ in SAC
}
    \label{fig:7}
\end{figure}

To optimize the performance of SAC, we conducted separate tests on the threshold $\sigma$ selection and data sliding hyperparameter $k$ settings and their impact on the overall performance, as shown in Fig.~\ref{fig:7}. 

Fig.~\ref{fig:7} (a) illustrates the relationship between the threshold and TSNet-SAC performance, where the utility index is indicated by the color temperature, with lower color temperatures indicating better performance. 

It can be observed that the threshold is dynamic fluctuating with the access count, especially, the optimal value is around $0.3$ with a slow decreasing trend. 

Fig.~\ref{fig:7} (b) illustrates the gradient of gain caused by hyperparameter k on the system performance. It can be seen that although an increasing $k$ can endlessly improve performance, there is a clear marginal effect of it, i.e., there exists a trade-off $k$ around $20$ with an abrupt change in gradient, which balances the improvement of network performance and computing costs. In the actual deployment of TSNet-SAC, adaptive control parameters and thresholds are adjusted to achieve the best overall performance.
 
\section{Conclusion}\label{Conclusion}

We proposed the TSNet, which employs the Transformer architecture and is trained using a dataset generated by the GA for task offloading and resource allocation. Furthermore, we proposed the Extender component to enhance network scalability. To address the inherent limitations of the Genetic Algorithm (GA), we designed the Sliding Augment Component (SAC) to improve network performance, robustness, and scalability. Compared with traditional algorithms, TSNet-SAC exhibited higher efficiency in real-time task scheduling. In light of the current research limitations, this paper aims to propose a novel paradigm for task scheduling, in the hope of supplying the ongoing discourse with further exploration.

\vspace{12pt}
\color{red}
\end{document}